\documentclass[sigconf]{acmart}
\AtBeginDocument{%
  }

\setcopyright{acmlicensed}
\copyrightyear{2018}
\acmYear{2018}
\acmDOI{XXXXXXX.XXXXXXX}
\acmISBN{978-1-4503-XXXX-X/2018/06}

\acmSubmissionID{8191}

\usepackage{amsthm}
\theoremstyle{remark}
\newtheorem{remark}{Remark}[section]

\usepackage{booktabs}
\usepackage{multirow}
\usepackage{makecell}
\usepackage[table]{xcolor}

\begin{document}

\title{Scaling Exposes the Trigger: Input-Level Backdoor Detection in Text-to-Image Diffusion Models via Cross-Attention Scaling}

\author{Zida Li}
\authornote{Both authors contributed equally to this research.}
\email{202463020004@nuist.edu.cn}
\author{Jun Li}
\authornotemark[1]
\email{lijuun@yeah.net}
\affiliation{%
  \institution{Nanjing University of Information Science and Technology}
  \city{Nanjing}
  \country{China}
}

\author{Yuzhe Sha}
\email{202383290260@nuist.edu.cn}
\author{Ziqiang Li}
\authornote{corresponding author.}
\email{iceli@mail.ustc.edu.cn}
\affiliation{%
  \institution{Nanjing University of Information Science and Technology}
  \city{Nanjing}
  \country{China}
}

\author{Lizhi Xiong}
\email{lzxiong16@163.com}
\author{Zhangjie Fu}
\email{fzj@nuist.edu.cn}
\affiliation{%
  \institution{Nanjing University of Information Science and Technology}
  \city{Nanjing}
  \country{China}
}

\renewcommand{\shortauthors}{Li et al.}

\begin{abstract}
Text-to-image (T2I) diffusion models have achieved remarkable success in image synthesis, but their reliance on large-scale data and open ecosystems introduces serious backdoor security risks. Existing defenses, particularly input-level methods, are more practical for deployment but often rely on observable anomalies that become unreliable under stealthy, semantics-preserving trigger designs. As modern backdoor attacks increasingly embed triggers into natural inputs, these methods degrade substantially, raising a critical question: can more stable, implicit, and trigger-agnostic differences between benign and backdoor inputs be exploited for detection?
In this work, we address this challenge from an active probing perspective. We introduce controlled scaling perturbations on cross-attention and uncover a novel phenomenon termed \textbf{Cross-Attention Scaling Response Divergence (CSRD)}, where benign and backdoor inputs exhibit systematically different response evolution patterns across denoising steps. Building on this insight, we propose \textbf{SET}, an input-level backdoor detection framework that constructs response-offset features under multi-scale perturbations and learns a compact benign response space from a small set of clean samples. Detection is then performed by measuring deviations from this learned space, without requiring prior knowledge of the attack or access to model training.
Extensive experiments demonstrate that SET consistently outperforms existing baselines across diverse attack methods, trigger types, and model settings, with particularly strong gains under stealthy implicit-trigger scenarios. Overall, SET improves AUROC by 9.1\% and ACC by 6.5\% over the best baseline, highlighting its effectiveness and robustness for practical deployment.

\end{abstract}







\maketitle

\section{Introduction}

\begin{figure}[t]
  \centering
  \includegraphics[width=\columnwidth]{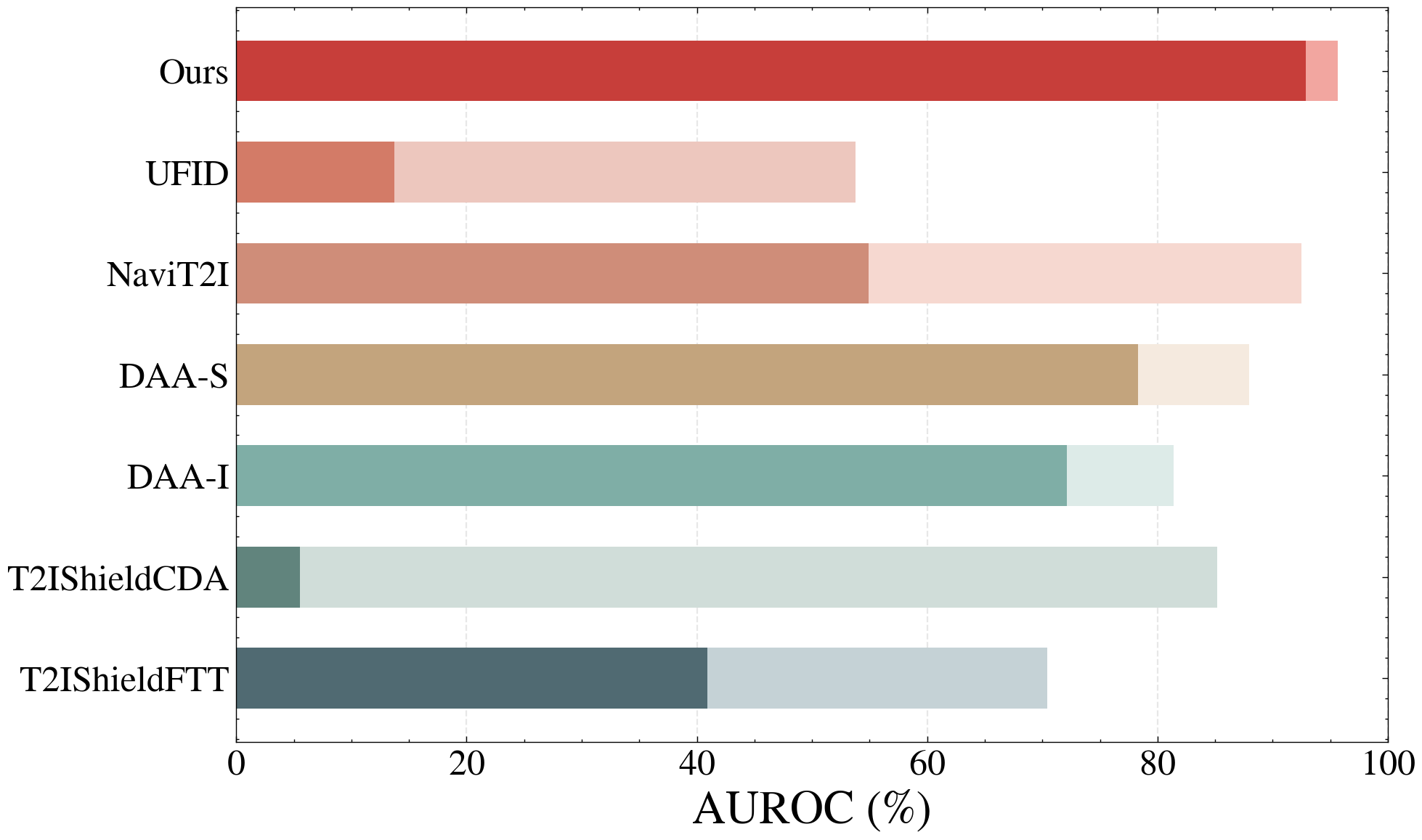}
  \caption{Robustness to stealthy and explicit triggers. Dark bars denote the AUROC under the stealthy IBA attack, while light bars represent the deviation from the mean AUROC across four explicit-trigger attacks.}
  \label{fig:comparison}
\end{figure}

Recent years have witnessed the remarkable success of text-to-image (T2I) diffusion models~\cite{nichol2022glide,ramesh2022hierarchical,saharia2022imagen,rombach2022ldm} in image synthesis, leading to their widespread adoption in content creation~\cite{lyu2022humanai}, visual design~\cite{chong2025prompting}, and human-computer interaction~\cite{rapp2025genai}. Although these models achieve strong generation performance, they typically rely on large-scale text--image data and an open model ecosystem~\cite{truong2025diffusionsurvey,zhang2025t2isurvey}, which also introduce new security risks. In particular, when training data are collected from complex and insufficiently audited sources, adversaries may implant backdoors through data poisoning~\cite{gu2017badnets,zhai2023badt2i}, parameter manipulation~\cite{chou2023howtobackdoor,chen2023trojdiff,wang2024eviledit}, or partial fine-tuning~\cite{huang2024personalization}. A poisoned model can exhibit normal behavior on benign inputs while reliably generating attacker-specified harmful or NSFW content once the trigger is activated~\cite{li2024backdoorsurvey,struppek2023rickrolling,zhang2025iba}. Owing to their long-term stealth, low activation cost, and the naturalness and diversity of trigger patterns at inference time, such malicious behaviors pose substantial risks to generation safety, system reliability, and platform compliance. Consequently, developing effective defenses against backdoor attacks in text-to-image diffusion models has become an important and urgent research problem.

Existing defenses against backdoor attacks in text-to-image diffusion models can be broadly categorized into three groups: training-level~\cite{mo2024terd}, model-level~\cite{aravindan2025skdcag,jha2025sau}, and input-level methods~\cite{wang2024t2ishield,wang2026daa,zhai2025navi,guan2025ufid}. This taxonomy is also consistent with recent surveys on backdoor learning and diffusion-model security~\cite{li2024backdoorsurvey,truong2025diffusionsurvey,zhang2025t2isurvey}. Training-level approaches aim to prevent the model from learning backdoor behaviors during optimization. However, they typically require full access to the training data and training pipeline, making them impractical for third-party models with unknown provenance. Model-level methods operate post hoc by analyzing potentially compromised models, identifying candidate triggers, and mitigating malicious behaviors through techniques such as pruning, unlearning, or parameter editing~\cite{aravindan2025skdcag,jha2025sau}. Nevertheless, these approaches face significant challenges in text-to-image diffusion models due to the large trigger space and the complex interactions among multiple architectural components~\cite{truong2025diffusionsurvey,zhang2025t2isurvey}. Moreover, such interventions may inadvertently degrade the model’s original generation quality. In contrast, input-level methods focus on detecting and filtering suspicious inputs at inference time, making them more suitable for real-world deployment scenarios~\cite{wang2024t2ishield,wang2026daa,zhai2025navi,guan2025ufid}.
Despite their practicality, most existing input-level methods rely on two types of observable anomalous signals during inference: \textit{(i) surface-level anomalies directly exposed under standard generation}, and \textit{(ii) abnormal responses indirectly revealed under generic perturbations}~\cite{wang2024t2ishield,wang2026daa,zhai2025navi,guan2025ufid}. This reliance becomes increasingly problematic as trigger designs evolve from explicit and rare anomalous tokens~\cite{struppek2023rickrolling} to more natural, semantics-preserving sentence-level triggers embedded within full inputs~\cite{huang2024personalization,wang2024eviledit,zhang2025iba}. Recent stealthy backdoor attacks, such as IBA~\cite{zhang2025iba}, exemplify this trend. Rather than depending on isolated trigger tokens, they embed the trigger mechanism into natural user inputs using semantically consistent and syntactically plausible expressions, often combined with additional constraints. Consequently, these attacks can maintain high attack success rates while significantly reducing the visibility of anomalous signals, both in internal model responses and at the semantic level~\cite{zhang2025iba}. As illustrated in Fig.~\ref{fig:comparison}, the effectiveness of existing input-level defenses deteriorates substantially under such stealthy, implicit-trigger settings. This observation raises a critical question: \textbf{when surface anomalies become indistinguishable under stealthy triggering conditions, do backdoor and benign inputs still exhibit stable, implicit, and trigger-agnostic differences that can be reliably exploited for detection?}

\begin{figure*}[t]
    \centering
    \includegraphics[width=\textwidth]{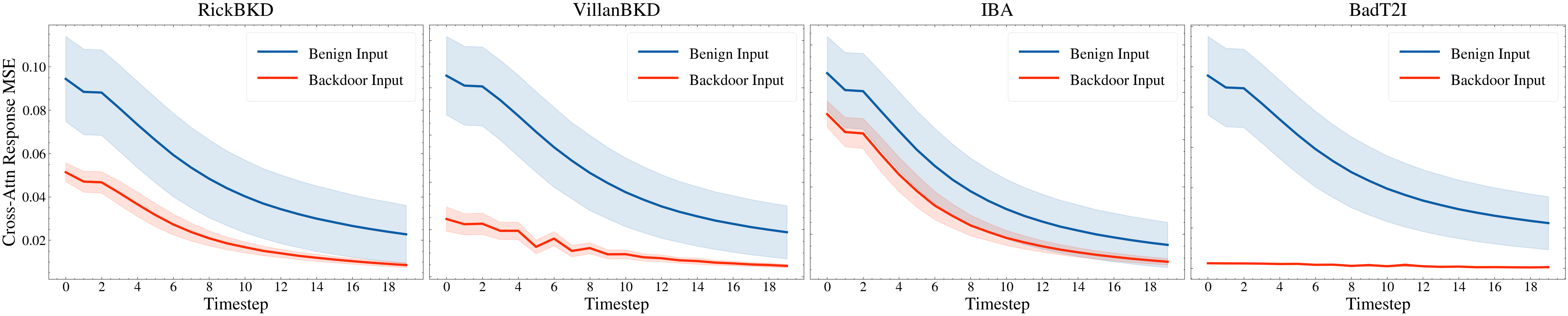}
    \caption{\textbf{Cross-Attention Scaling Response Divergence.} We measure the response difference of an input along the conditional injection pathway by computing the mean squared error between the scaled response and the reference response in four representative backdoored T2I models: RickBKD, VillanBKD, IBA, and BadT2I. As shown in the figure, after scaling, benign and backdoor inputs exhibit stable and separable response shifts.}
    \label{fig:concat}
\end{figure*}

To investigate this question, we adopt an active probing perspective for input-level backdoor detection. The central challenge lies in designing a perturbation mechanism that can reliably expose the subtle and implicit discrepancies induced by stealthy backdoors, without relying on explicit trigger patterns or conspicuous surface anomalies. This necessitates a probing strategy that is intrinsically aligned with the conditional generation process and capable of amplifying latent differences between benign and backdoor inputs. Motivated by prior active probing approaches such as SCALE-UP~\cite{guo2023scaleup} and IBD-PSC~\cite{hou2024ibdpsc}, we introduce controlled scaling perturbations on the cross-attention mechanism and analyze the resulting evolution of internal model responses. Preliminary experiments reveal that benign and backdoor inputs exhibit qualitatively distinct trajectories in cross-attention response MSE across denoising steps $t$. Importantly, this separation persists across diverse backdoor settings, as illustrated in Fig.~\ref{fig:concat}. Further comprehensive analysis confirms that such probing induces stably separable response evolution patterns. Although the specific direction of divergence may vary depending on the attack configuration, the discrepancy remains consistent and systematic. We refer to this phenomenon as \textbf{Cross-Attention Scaling Response Divergence (CSRD)}.

Based on this observation, we propose \textbf{SET}, a simple yet effective input-level backdoor detection framework for text-to-image diffusion models. Given a suspicious deployed model, SET applies controlled scaling to cross-attention scores and tracks the resulting variations in cross-attention responses throughout the denoising process. Using the perturbation-induced response variations of a small set of benign samples under different scaling conditions and across cross-attention layers, SET constructs a benign response space that captures the stable benign response structure under active probing. Moreover, instead of relying on a manually fixed threshold, SET adaptively learns the boundary of the benign sample space with a soft-boundary one-class objective~\cite{scholkopf2001estimating,ruff2018deep}, yielding a more compact and flexible benign region in the embedding space and improving robustness across diverse attack scenarios. During inference, backdoor detection is performed by measuring the deviation of a test input from this benign response space.
Extensive experiments show that SET consistently outperforms existing baselines across diverse attack methods, trigger types, model configurations, and adaptive settings, with especially strong gains under stealthy implicit-trigger attacks where existing methods often degrade. Overall, SET improves AUROC by an average of 9.1\% and ACC by 6.5\% over the best-performing baseline. These results demonstrate the effectiveness of controlled cross-attention scaling for practical input-level backdoor detection in text-to-image diffusion models.
In summary, our main contributions are as follows:
\begin{itemize}
    \item We identify a simple yet important phenomenon, termed \textbf{Cross-Attention Scaling Response Divergence (CSRD)}: scaling cross-attention responses along the text-conditional injection pathway induces response divergence between benign and backdoor inputs. We further provide a theoretical analysis to explain the origin of this phenomenon.
    
    \item Based on this observation, we propose \textbf{SET}, a simple yet effective input-level backdoor detector for text-to-image diffusion models. SET constructs response-offset features through cross-attention scaling and learns a compact benign response space from a small set of benign samples, enabling effective detection of suspicious inputs. The method requires no prior knowledge of the attack, and its decision boundary can adapt automatically to the target poisoned model, making it practical for real-world deployment.

    % \item We conduct extensive experiments on benchmark datasets and systematically evaluate SET under diverse attack methods and trigger types. Experimental results show that SET consistently outperforms existing baselines across a broader range of attack settings, improving average AUROC and ACC by 9.1\% and 6.5\%, respectively, over the best baseline. These results confirm the effectiveness and robustness of the proposed method.
    \item Extensive experiments on benchmark datasets show that SET consistently outperforms existing baselines under diverse attack methods and trigger types. Across a wide range of attack settings, SET achieves average gains of 9.1\% in AUROC and 6.5\% in ACC over the best baseline. These results validate the effectiveness and robustness of SET.
\end{itemize}

\section{Related Works}
\subsection{Text-to-Image Diffusion Models}
% Text-to-image diffusion models generate images from noise by injecting textual conditions into an iterative denoising process~\cite{ho2020ddpm}. Early works such as GLIDE~\cite{nichol2022glide}, DALL$\cdot$E 2~\cite{ramesh2022hierarchical}, and Imagen~\cite{saharia2022imagen} demonstrated the effectiveness of diffusion-based text-conditioned generation. Latent Diffusion Models~\cite{rombach2022ldm} further improved efficiency by moving denoising to the latent space and established cross-attention~\cite{vaswani2017attention} as the standard mechanism for text injection, typically in conjunction with powerful text encoders such as CLIP~\cite{radford2021clip}. Building on these foundations, subsequent studies advanced controllable and personalized generation, as represented by ControlNet~\cite{zhang2023controlnet}, Textual Inversion~\cite{gal2023textualinversion}, and DreamBooth~\cite{ruiz2023dreambooth}. With the open release of models such as Stable Diffusion~\cite{rombach2022ldm}, T2I diffusion models have rapidly evolved into an open ecosystem for model sharing and adaptation~\cite{truong2025diffusionsurvey,zhang2025t2isurvey}, accelerating real-world adoption while also increasing the risks of tampering, poisoning, and backdoor injection.

Text-to-image diffusion models generate images from noise by injecting textual conditions into an iterative denoising process~\cite{ho2020ddpm}. Early works such as GLIDE~\cite{nichol2022glide}, DALL$\cdot$E 2~\cite{ramesh2022hierarchical}, and Imagen~\cite{saharia2022imagen} demonstrated the effectiveness of diffusion-based text-conditioned image generation. Latent Diffusion Models~\cite{rombach2022ldm} further improved overall efficiency by moving denoising to the latent space and established cross-attention~\cite{vaswani2017attention} as the standard mechanism for text injection, typically in conjunction with powerful pretrained text encoders such as CLIP~\cite{radford2021clip}. Building on these foundations, subsequent studies further advanced controllable and personalized generation, as represented by ControlNet~\cite{zhang2023controlnet}, Textual Inversion~\cite{gal2023textualinversion}, and DreamBooth~\cite{ruiz2023dreambooth}. With the open release of models such as Stable Diffusion~\cite{rombach2022ldm}, T2I diffusion models have rapidly evolved into an open ecosystem for model sharing and adaptation~\cite{truong2025diffusionsurvey,zhang2025t2isurvey}, accelerating real-world adoption while also increasing the risks of tampering, poisoning, and backdoor injection.

\subsection{Backdoor Attack on T2I Diffusion Models}

% Backdoor attacks aim to preserve normal model behavior on benign inputs while inducing attacker-specified outputs under trigger conditions~\cite{gu2017badnets,li2024backdoorsurvey}. In text-to-image (T2I) diffusion models, this threat extends beyond label manipulation to fine-grained control over generated objects, attributes, styles, and even stealthy semantic content~\cite{zhai2023badt2i,huang2024personalization,wang2024eviledit,zhang2025iba}, drawing increasing attention to backdoor vulnerabilities in conditional generation~\cite{chou2023howtobackdoor,chen2023trojdiff,chou2023villandiffusion}.
% Early studies on T2I backdoor attacks~\cite{struppek2023rickrolling,chou2023villandiffusion,zhai2023badt2i} primarily focused on attack feasibility and injection pathways. Rickrolling the Artist~\cite{struppek2023rickrolling} demonstrated that the text encoder can serve as a practical entry point for backdoor injection via visually similar character triggers. VillanDiffusion~\cite{chou2023villandiffusion} extended backdoor attacks to diffusion models in both conditional and unconditional settings, while BadT2I~\cite{zhai2023badt2i} showed that large-scale T2I models such as Stable Diffusion can be effectively poisoned with only limited fine-tuning.

Backdoor attacks aim to preserve normal model behavior on benign inputs while inducing attacker-specified malicious outputs under specific trigger conditions~\cite{gu2017badnets,li2024backdoorsurvey}. In text-to-image (T2I) diffusion models, this threat extends far beyond simple label manipulation to fine-grained control over generated objects, visual attributes, artistic styles, and even stealthy semantic content~\cite{zhai2023badt2i,huang2024personalization,wang2024eviledit,zhang2025iba}, thereby drawing increasing research attention to backdoor vulnerabilities in conditional generative models~\cite{chou2023howtobackdoor,chen2023trojdiff,chou2023villandiffusion}.
Early studies on T2I backdoor attacks~\cite{struppek2023rickrolling,chou2023villandiffusion,zhai2023badt2i} primarily focused on attack feasibility, threat modeling, and backdoor injection pathways. Rickrolling the Artist~\cite{struppek2023rickrolling} demonstrated that the text encoder can serve as a practical and effective entry point for backdoor injection via visually similar character triggers. VillanDiffusion~\cite{chou2023villandiffusion} further extended backdoor attacks to diffusion models in both conditional and unconditional generation settings, while BadT2I~\cite{zhai2023badt2i} showed that large-scale T2I models such as Stable Diffusion can be effectively poisoned through only limited additional fine-tuning.

More recent work has shifted toward reducing attack cost and improving stealth. Personalization as a Shortcut~\cite{huang2024personalization} exploits personalization pipelines for efficient few-shot backdoor injection, and EvilEdit~\cite{wang2024eviledit} further explores training-free and data-free model editing. Meanwhile, IBA~\cite{zhang2025iba} enhances stealth by minimizing anomalous traces in both semantic consistency and internal model responses while maintaining a high attack success rate. Overall, T2I backdoor attacks are evolving toward more diverse entry points, lower cost, and increasingly concealed malicious behaviors, posing growing challenges for practical deployment-time defenses, particularly input-level methods~\cite{li2024backdoorsurvey,zhang2025t2isurvey}.

% More recent work has shifted toward reducing attack cost and improving stealth. Personalization as a Shortcut~\cite{huang2024personalization} exploits personalization pipelines for few-shot backdoor injection, and EvilEdit~\cite{wang2024eviledit} further explores training-free and data-free model editing. Meanwhile, IBA~\cite{zhang2025iba} enhances stealth by minimizing anomalous traces in both semantic consistency and internal responses while maintaining a high attack success rate. Overall, T2I backdoor attacks are evolving toward more diverse entry points, lower cost, and increasingly concealed malicious behaviors, posing growing challenges for deployment-time defenses, particularly input-level methods~\cite{li2024backdoorsurvey,zhang2025t2isurvey}.

\subsection{Backdoor Defense on T2I Diffusion Models}
% Existing defenses against backdoor attacks on T2I diffusion models can be grouped into three categories: training-level, model-level, and input-level methods~\cite{li2024backdoorsurvey,truong2025diffusionsurvey,zhang2025t2isurvey}. These categories differ in their defense capabilities, resource requirements, and deployment settings.
Existing defenses against backdoor attacks on T2I diffusion models can be broadly categorized into three main types: training-level, model-level, and input-level methods~\cite{li2024backdoorsurvey,truong2025diffusionsurvey,zhang2025t2isurvey}. These categories differ substantially with respect to defense effectiveness, computational and resource requirements, and applicability across different practical deployment scenarios.

Training-level methods are applied during data construction or optimization and aim to prevent the model from learning backdoor features through data filtering, robust training, regularization, or conditional perturbation~\cite{mo2024terd}. They can reduce risk at the source, but they require access to the training data and pipeline. Model-level methods, by contrast, repair suspicious pretrained models by weakening the association between triggers and target outputs, typically through parameter editing, concept erasure, local forgetting, or distillation. Representative methods such as SKD-CAG~\cite{aravindan2025skdcag} and SAU~\cite{jha2025sau} follow this line, but these approaches generally require access to model parameters or internal representations and may incur additional repair cost.

Input-level methods focus on deployment-time detection without modifying model parameters, which makes them more suitable for third-party model use and online safety filtering~\cite{wang2024t2ishield,wang2026daa,zhai2025navi,guan2025ufid}. Existing methods in this category can be further divided into two groups. The first group is based on text perturbation or rewriting. These methods disrupt potential triggers through synonym substitution, word-order changes, or semantic rewriting, and then infer maliciousness from whether the attack effect is weakened. They are simple and model-agnostic, but their effectiveness often declines when triggers become more natural and more tightly coupled with benign semantics~\cite{zhang2025t2isurvey}.

The second group analyzes internal model responses and looks for observable traces left by backdoor triggers during conditional injection and iterative denoising. T2IShield~\cite{wang2024t2ishield} studies semantic assimilation in cross-attention and argues that trigger tokens abnormally absorb the attention patterns of other tokens, which disrupts the normal division of semantic roles. Based on this observation, it proposes two detection methods. FTT measures the global structural contraction of attention maps with the Frobenius norm and performs coarse-grained detection through threshold truncation. CDA characterizes fine-grained structural correlations among attention maps using covariance matrices and conducts discriminant analysis on a Riemannian manifold to capture subtler structural shifts. DAA~\cite{wang2026daa} further examines the dynamic evolution of cross-attention during denoising and shows that backdoor inputs consistently exhibit temporal imbalance in the attention assigned to the $\langle \mathrm{EOS} \rangle$ token. It therefore introduces two variants, DAA-I and DAA-S. The former treats the attention map of each token as spatially independent and extracts dynamic features with the Frobenius norm, while the latter additionally models spatial correlations and temporal propagation among attention maps through a graph state equation. NaviT2I~\cite{zhai2025navi} detects backdoor inputs by exploiting abnormal neural activation changes induced by trigger tokens in the early diffusion stage. Specifically, it measures early activation differences through token-wise masking and combines them with semantic-distance normalization to obtain token-level scores. It then constructs a detection statistic based on the outlierness of abnormal tokens and sets the threshold by fitting the distribution of clean samples. In black-box settings, UFID~\cite{guan2025ufid} identifies malicious inputs by comparing the consistency of generated results under input perturbations. Concretely, it constructs perturbed variants of the same input, generates multiple outputs, builds a weighted similarity graph, computes a graph-density score, and determines the detection threshold using a small set of clean validation samples.

Overall, existing defenses have substantially improved backdoor mitigation for T2I diffusion models, covering training-time prevention, post-hoc model repair, and deployment-time input screening~\cite{li2024backdoorsurvey,truong2025diffusionsurvey,zhang2025t2isurvey}. Recent input-level methods, in particular, have expanded the range of detection signals from surface semantics to internal attention, activation, and generation-consistency patterns~\cite{wang2024t2ishield,wang2026daa,zhai2025navi,guan2025ufid}. However, many current methods still depend on anomalous cues that are easier to expose under standard inference or generic perturbations. As triggers become more natural, stealthier, and more semantically aligned with benign prompts, these cues often become less stable, which limits robustness across attack settings~\cite{wang2024eviledit,zhang2025iba}. Some methods also require multiple model queries or additional threshold calibration, which may reduce their practical deployability~\cite{guan2025ufid}.

\section{Cross-Attention Scaling Response Divergence}\label{sec:3}
% In text-to-image diffusion models, textual conditions are incorporated into the denoising process primarily via cross-attention mechanisms~\cite{vaswani2017attention,rombach2022ldm}. When this conditional pathway is perturbed through controlled scaling, the model exhibits systematic variations in its internal responses. Our empirical analysis reveals that benign and backdoor inputs follow distinctly different response patterns under such perturbations. Specifically, across multiple cross-attention layers and a range of scaling factors, these inputs yield consistently divergent response trajectories. We term this phenomenon \textbf{Cross-Attention Scaling Response Divergence (CSRD)}. Notably, CSRD does not rely on prior knowledge of specific attack mechanisms; instead, it characterizes intrinsic differences in how inputs propagate through the conditional pathway, thereby providing a principled foundation for the detection method proposed in this work.

In text-to-image diffusion models, textual conditions are incorporated into the denoising process primarily via cross-attention mechanisms~\cite{vaswani2017attention,rombach2022ldm}. When the attention scores are perturbed through controlled scaling, the induced change first appears in the attention distribution and is subsequently propagated to the resulting cross-attention output. Our empirical analysis reveals that benign and backdoor inputs differ specifically in how this scaling operation changes the resulting cross-attention outputs relative to the unscaled case: benign inputs typically induce relatively small and smooth output shifts, whereas backdoor inputs produce larger and often less regular deviations. This discrepancy persists across multiple cross-attention layers and scaling factors, leading to consistently different response trajectories. We term this phenomenon \textbf{Cross-Attention Scaling Response Divergence (CSRD)}.

This section first reviews the cross-attention mechanism in text-to-image diffusion models and its role in injecting textual conditions into the denoising process. We then characterize the proposed CSRD phenomenon under controlled cross-attention scaling, showing that benign and backdoor inputs exhibit systematically different response evolution patterns. Finally, we present a theoretical analysis that explains the origin of this divergence.

Textual conditions in text-to-image diffusion models are typically injected into the UNet denoising process via cross-attention~\cite{vaswani2017attention,rombach2022ldm}. Given a text prompt $x$, a text encoder (\textit{e.g.}, CLIP~\cite{radford2021clip}) maps it to a contextual embedding $E(x) \in \mathbb{R}^{m \times d}$, where $m$ denotes the number of text tokens and $d$ is the embedding dimension. At denoising step $t$ and the $\ell$-th cross-attention layer, let $H_{t,\ell} \in \mathbb{R}^{n \times d}$ denote the visual feature, where $n$ is the number of spatial positions. The corresponding query, key, and value matrices are then defined as:
\begin{equation}
Q_{t,\ell} = H_{t,\ell} W_Q, \qquad
K_{t,\ell} = E(x) W_K, \qquad
V_{t,\ell} = E(x) W_V,
\end{equation}
where $W_Q$, $W_K$, and $W_V$ are learnable projection matrices. The attention score matrix is then computed as:
\begin{equation}
S_{t,\ell} = \frac{Q_{t,\ell} K_{t,\ell}^{\top}}{\sqrt{d}}.
\end{equation}
Accordingly, the standard cross-attention response is given by:
\begin{equation}
R_{t,\ell} = \mathrm{softmax}(S_{t,\ell}) V_{t,\ell}.
\end{equation}
In this paper, we introduce a controlled scaling operation on the attention scores prior to response computation. Specifically, the scaled cross-attention response is defined as:
\begin{equation}
R_{t,\ell}^{\lambda}(x) = \mathrm{softmax}(\lambda S_{t,\ell}) V_{t,\ell},
\end{equation}
where $\lambda$ denotes the scaling factor, and $\lambda = 1$ reduces to the standard cross-attention response. The scaled response is then fused with the current visual feature and propagated through the subsequent residual blocks and denoising updates.
In latent-diffusion-based text-to-image models, cross-attention serves as the primary mechanism through which textual conditions are injected into the latent representation~\cite{rombach2022ldm}. For benign inputs, the conditional response follows the normal semantic composition process, such that scaling the attention scores perturbs the response in a manner that remains consistent with standard text-guided generation. In contrast, for backdoor inputs, the poisoned model exhibits a different response under the same textual pattern, because the implanted backdoor alters how textual conditions are translated into latent guidance. Consequently, controlled scaling of the cross-attention scores drives benign and backdoor inputs along distinct conditional response trajectories, resulting in systematically different response shifts across layers and denoising steps.

%The scaled response is then fused with the current visual feature and passed through the subsequent residual blocks and denoising updates. Cross-attention thus provides the main mechanism by which textual conditions enter the latent representation in latent-diffusion-based T2I models~\cite{rombach2022ldm}. For benign inputs, the conditional response follows normal semantic composition, so scaling the attention scores changes the response in a way that remains consistent with standard text-guided generation. For backdoored inputs, the poisoned model responds differently under the same textual input pattern, because the learned backdoor changes how textual conditions are mapped to latent guidance. As a result, scaling the cross-attention scores causes benign and backdoored inputs to diverge from the same conditional response trajectory, leading to distinct response shifts across layers and denoising steps.

\noindent\textbf{Settings.} To illustrate this phenomenon, we consider two representative backdoor attacks, Rickrolling~\cite{struppek2023rickrolling} and IBA~\cite{zhang2025iba}. Rickrolling corresponds to a relatively explicit trigger mechanism, whereas IBA represents a more stealthy attack scenario. For each attack, we compare benign and backdoor inputs under identical generation settings and analyze how their responses vary with the scaling factor, denoising step, and cross-attention layer. 

\noindent\textbf{Results.} As shown in Fig.~\ref{fig:concat}, benign and backdoor inputs exhibit clearly distinct response trajectories under cross-attention scaling in the compromised diffusion model. Although Rickrolling and IBA differ substantially in trigger form and degree of stealthiness, both produce stable and separable response shifts. The key characteristic of CSRD is that benign and backdoor inputs follow systematically different response patterns under scaling perturbations. This regularity is more robust than a discrepancy observed at a single operating point, and therefore provides a stronger foundation for subsequent detection.

% To show that CSRD is not merely an incidental effect, we next present both theoretical and empirical analyses.
    
\begin{theorem}
\label{thm:local_separation}
Fix a denoising step $t$ and a cross-attention layer $\ell$.
For each input $x$ and scaling factor $\lambda$, define the scaling-induced response shift as:
\begin{equation}
D_{t,\ell}(x;\lambda)
:=
\operatorname{MSE}\!\left(R_{t,\ell}^{\lambda}(x),\,R_{t,\ell}^{1}(x)\right).
\end{equation}
For each class $c \in \{\mathrm{ben}, \mathrm{bd}\}$, define the class-wise mean response-shift curve as:
\begin{equation}
\bar D_c^{t,\ell}(\lambda)
:=
\mathbb{E}_{x\sim P_c}\!\left[D_{t,\ell}(x;\lambda)\right].
\end{equation}
where $\mathrm{ben}$ and $\mathrm{bd}$ denote the benign and backdoor classes, respectively.

Then there exist class-dependent coefficients $\Gamma_{\mathrm{ben}}^{t,\ell}$ and $\Gamma_{\mathrm{bd}}^{t,\ell}$ such that, for each $c \in \{\mathrm{ben}, \mathrm{bd}\}$,
\begin{equation}
\bar D_c^{t,\ell}(\lambda)
=
\Gamma_c^{t,\ell}(\lambda-1)^2
+
o\!\left((\lambda-1)^2\right)
\qquad
\text{as }\lambda\to 1.
\end{equation}

In particular, if
\begin{equation}
\Gamma_{\mathrm{ben}}^{t,\ell}\neq \Gamma_{\mathrm{bd}}^{t,\ell},
\end{equation}
then there exists $\epsilon>0$ such that
\begin{equation}
\bar D_{\mathrm{bd}}^{t,\ell}(\lambda)\neq \bar D_{\mathrm{ben}}^{t,\ell}(\lambda),
\qquad
\forall\,\lambda \text{ satisfying } 0<|\lambda-1|<\epsilon.
\end{equation}
That is, benign and backdoor inputs are locally separable under cross-attention scaling at $(t,\ell)$.

Moreover, the sign of
\begin{equation}
\bar D_{\mathrm{bd}}^{t,\ell}(\lambda)-\bar D_{\mathrm{ben}}^{t,\ell}(\lambda)
\end{equation}
depends on the attack and is not fixed a priori.
\end{theorem}

\begin{remark}
Theorem~\ref{thm:local_separation} directly motivates the design of our method. It shows that, around $\lambda=1$, benign and backdoor inputs exhibit different second-order response sensitivities under cross-attention scaling, which leads to local separability in the class-wise response-shift curves. This suggests that the discriminative information for backdoor detection is not confined to a single isolated scaling point, but instead lies in a stable discrepancy in how the two classes respond to scaling perturbations. Motivated by this observation, we construct detection features by aggregating response shifts over multiple nearby scaling factors, so as to capture this intrinsic response-pattern difference more effectively. The proof is provided in \textbf{Appendix A}.
\end{remark}

\section{Method}

\subsection{Threat Model}

We consider a practical deployment scenario involving two roles: the attacker, who implants a backdoor into a text-to-image (T2I) model before release, and the defender, who adopts the suspicious third-party model for downstream applications and aims to detect backdoor inputs at inference time.

\noindent\textbf{Attacker.}
The attacker implants a backdoor into a text-to-image (T2I) model during model development or adaptation and releases the compromised model under the guise of a benign pre-trained model. The attack is designed such that the model behaves normally on benign inputs but generates attacker-specified target content when presented with backdoor-triggering inputs at inference time.

\noindent\textbf{Defender.}
The defender adopts the suspicious T2I model for downstream applications and aims to determine, at inference time, whether a given input is benign or backdoored. We assume that the defender has white-box access to the model parameters and intermediate responses, which enables the extraction of internal detection features. However, the defender has no prior knowledge of the attack, including the training data, poisoning strategy, trigger patterns, or target concepts. We further assume that only a small benign reference set is available. Under these constraints, the defender seeks to perform input-level backdoor detection without modifying the model or relying on attack-specific prior knowledge.

% \subsection{Preliminaries}

% \textbf{Threat model.} We consider a deployment scenario in which an untrusted third-party text-to-image (T2I) model is adopted for downstream applications. In this setting, an attacker may implant a backdoor into the T2I model and release the compromised model under the guise of a benign pre-trained model. The defender is assumed to have white-box access to the model parameters and intermediate responses, which enables the extraction of detection features, but has no prior knowledge of the attack, including the training data, backdoor triggers, or target concepts. We further assume that only a small benign reference set is available to the defender.

% \noindent\textbf{Defender Goals.} An ideal input-level backdoor detector should accurately identify backdoored samples while remaining practical for real-world deployment. Motivated by this objective, we ask whether the model's internal responses contain stable diagnostic signals that can be extracted using only a small benign reference set, without relying on prior knowledge of the attack, such as trigger patterns or target outputs, thereby enabling reliable discrimination between benign and backdoored inputs.

\subsection{The Overview of SET}
\begin{figure}[t]
    \centering
    \includegraphics[width=1\linewidth]{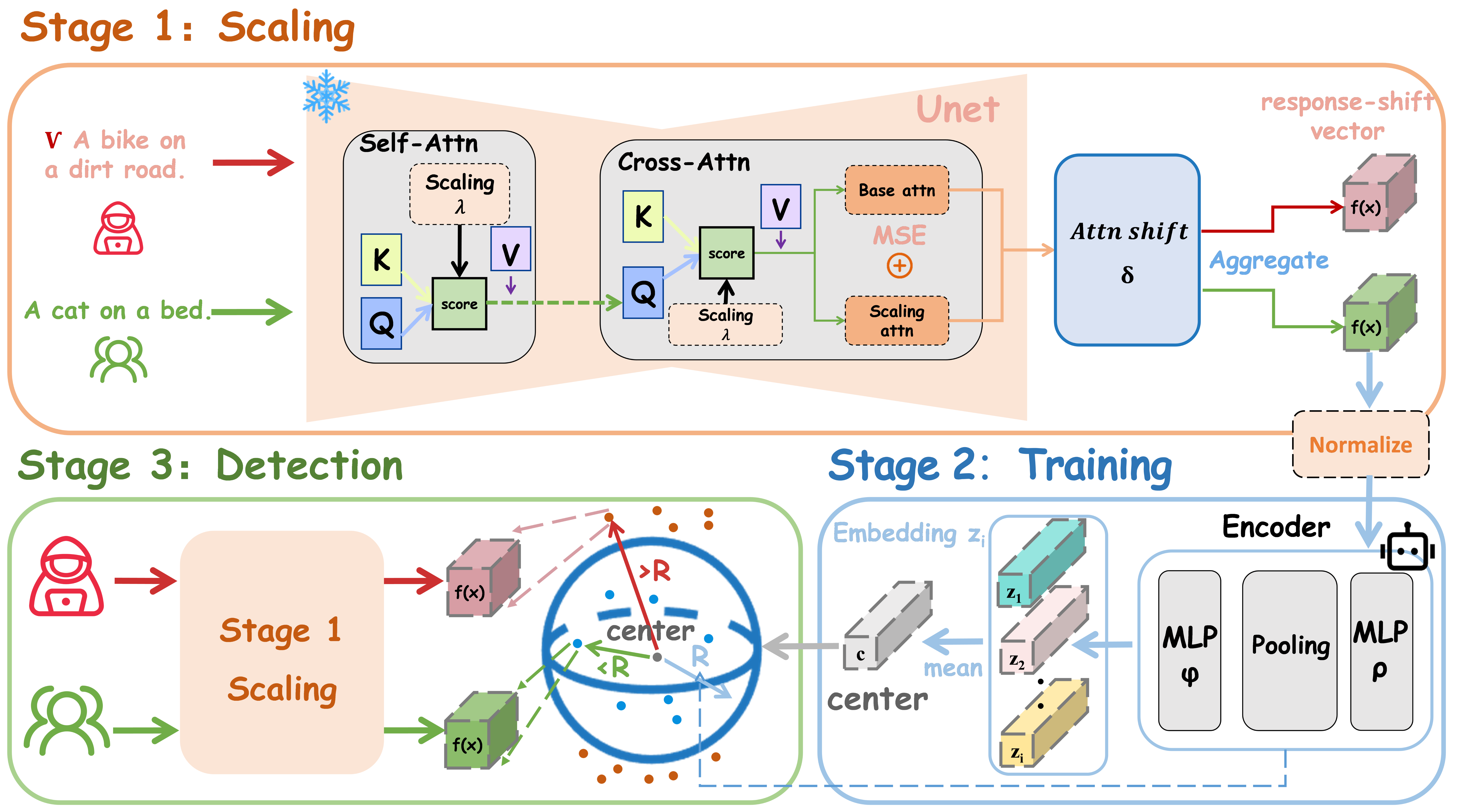}
    \caption{Overview of SET. We apply controlled scaling to selected self- and cross-attention modules in the U-Net to extract attention-response shifts and aggregate them into a response-shift vector $f(x)$. Benign vectors are normalized and encoded into low-dimensional embeddings to build a compact benign space by estimating a robust center $c$ and optimizing decision radius $R$ with a one-class objective. At test time, backdoor inputs are identified by their distance to the benign center.}
    \label{fig:set_overview}
\end{figure}

As shown in Sec.~\ref{sec:3}, scaling cross-attention produces stable and separable response differences for benign and backdoor inputs. SET builds on this observation to identify backdoor inputs. The method measures scaling-induced response shifts in selected attention modules and uses them as detection features.

An overview of the proposed SET framework is shown in Fig.~\ref{fig:set_overview}. SET has three stages: (1) scaling cross-attention responses, (2) learning a benign sample space, and (3) input detection. In the first stage, for a given input, we apply multiple scaling factors to selected cross-attention layers, measure the mean squared error between each scaled response and the reference response, and aggregate these shifts across scaling factors, layers, and denoising steps into a scaling response shift vector. In the second stage, we pass the standardized vector through a lightweight encoder and use a small set of benign samples to learn a compact benign space and an adaptive radius in the embedding space, enabling finer-grained detection. In the third stage, a test input goes through the same feature extraction and encoding pipeline, and the final decision is based on its distance from the benign center.

\subsection{Scaling Cross-Attention Responses}\label{sec:4.3}
Sec.~\ref{sec:3} shows that benign and backdoor inputs exhibit different response patterns under cross-attention scaling, which we refer to as CSRD. Building on this empirical observation, we use the resulting scaling-induced response shifts as discriminative features for subsequent detection.

Given a prompt $x$, we use a fixed scaling configuration with bidirectional scaling set $\Lambda$, selected denoising steps, and selected cross-attention layers. Specifically, we select cross-attention layers from the down and up blocks of the UNet, retaining those whose query length is $256$, corresponding to the $16\times16$ spatial resolution. We first run the model under the original setting, recording the reference cross-attention responses $R_{t,\ell}(x)$ for each selected denoising step $t$ and cross-attention layer $\ell$. We then apply scaling factors $\lambda\in\Lambda$ to the attention scores to obtain the corresponding scaled responses $R_{t,\ell}^{\lambda}(x)$. Although CSRD is defined on cross-attention, in practice we also apply the same scaling to self-attention scores during probing so the induced discrepancy propagates further through the latent feature space and becomes more pronounced. Detection features, however, are extracted only from cross-attention responses.

To quantify the effect of scaling, for each selected scaling factor, denoising step, and cross-attention layer, we compute the mean squared error between the scaled response and the reference response:
\begin{equation}
\delta_{t,\ell,\lambda}(x)=\operatorname{MSE}\!\left(R_{t,\ell}^{\lambda}(x),\,R_{t,\ell}(x)\right).
\end{equation}
In practice, this response difference is computed on the conditional branch. For each scaling factor $\lambda$, denoising step $t$, and cross-attention layer $\ell$, we measure the mean squared error between the scaled response and its reference response, and then average the squared difference over the value dimension, the attention heads, and the spatial positions to obtain a scalar feature for each $(\lambda, t, \ell)$. This scalar measures the magnitude of the perturbation-induced change in the cross-attention response under the given scaling condition and reflects how strongly the conditional injection pathway responds at a specific layer and denoising step. We then concatenate these scalar shifts across all selected scaling factors, denoising steps, and attention layers in a fixed order to form the scaling response shift vector $f(x)$ for input $x$. As a result, $f(x)$ captures the multi-scale, multi-step, and multi-layer response pattern of the input under controlled scaling, while remaining compact for efficient downstream modeling. In the next stage, this vector is normalized and used for benign sample space modeling and anomaly detection.

\subsection{Learning a Benign Sample Space}\label{sec:4.4}
After obtaining the scaling response shift vector defined in Sec.~\ref{sec:4.3},
we use a limited number of benign samples to learn a compact
benign sample space. Let the benign training set be $\mathcal{D}_b=\{x_i\}_{i=1}^{N}$. For each sample $x_i \in \mathcal{D}_b$, we first extract its scaling response shift vector $f(x_i)$ and perform dimension-wise standardization over the training set, yielding
\begin{equation}
\tilde{f}(x_i)=\frac{f(x_i)-\mu}{\sigma},
\end{equation}
where $\mu$ and $\sigma$ are the dimension-wise mean and standard deviation estimated from the benign training set. This standardization removes scale differences across dimensions, letting subsequent modeling focus on the shift pattern rather than raw magnitude. We then feed the standardized vector into a lightweight encoder $g_{\theta}$. The encoder first reshapes it into layer-wise response-shift tokens, applies a shared MLP to each token, aggregates the token features by their mean and standard deviation across layers, and maps the aggregated representation through a second MLP to obtain a low-dimensional embedding:
\begin{equation}
z_i = g_{\theta}\!\left(\tilde{f}(x_i)\right).
\end{equation}
In the embedding space, we first define a robust benign center $c$. After obtaining the embeddings of benign training samples, we compute their mean as a provisional center, discard a small fraction of samples with the largest distances to this provisional center, and then recompute $c$ as the mean of the remaining embeddings. In this way, $c$ captures the central tendency of benign response patterns in the embedding space. Once estimated, $c$ remains fixed during the subsequent training stage. We then use one-class learning to model the benign distribution by encouraging sample embeddings to cluster around this center, which defines a stable benign region. We adopt a soft-boundary one-class objective~\cite{scholkopf2001estimating,ruff2018deep} to optimize the encoder parameters $\theta$ and the decision radius $R$:
\begin{equation}
\min_{\theta,R}\; R^2+\frac{1}{\nu N }\sum_{i=1}^{N}\max\!\left(0,\left\|g_{\theta}\!\left(\tilde{f}(x_i)\right)-c\right\|_2^2-R^2\right),
\end{equation}
% where $R$ is the radius of the benign region, and $\nu$ controls the fraction of benign samples permitted to lie outside the boundary. This objective keeps most benign samples within a hypersphere centered at $c$ with radius $R$, while reducing the effect of a small number of outliers on the size of the benign region. Since the model is trained only on benign samples, this setting aligns with our threat model: the defender has no access to backdoor samples and no prior knowledge of the attack. Moreover, $R$ is learned adaptively from benign data rather than set by a manually chosen detection threshold.
where $R$ denotes the radius of the benign region, and $\nu$ specifies the fraction of benign samples allowed to fall outside the boundary. This objective constrains the majority of benign samples within a hypersphere centered at $c$ with radius $R$, while mitigating the influence of outliers on the size of the benign region. Since the model is trained exclusively on benign samples, this formulation is consistent with our threat model, in which the defender has no access to backdoor samples and no prior knowledge of the attack. Moreover, $R$ is estimated adaptively from benign data rather than determined by a manually specified detection threshold.

\subsection{Input Detection}
At test time, for any input prompt $x$ , we extract its scaling response shift vector and obtain its embedding representation $z(x)$ by following the procedure in Sections ~\ref{sec:4.3} and ~\ref{sec:4.4}. We then use the squared Euclidean distance to the benign center $c$ as the detection score:
\begin{equation}
s(x)=\|z(x)-c\|_2^2.
\end{equation}
This score measures how far the test input is from the benign sample space. If an input exhibits a response shift pattern under cross-attention scaling that is consistent with benign samples, its embedding typically lies close to the benign region and yields a small detection score. If its shift pattern departs substantially from the benign structure, its embedding lies farther from this region and produces a larger score.

Let $R$~denote the radius of the benign region learned in Section~\ref{sec:4.4}. The decision rule is:
\begin{equation}
\hat{y}(x)=
\begin{cases}
0, & s(x)\leq R^2,\\
1, & s(x)>R^2,
\end{cases}
\end{equation}
where $\hat{y}(x)=0$ indicates that the input is classified as benign, and $\hat{y}(x)=1$ indicates that the input is classified as backdoor.

% \begin{figure}
%     \centering
%     \includegraphics[width=1\linewidth]{kde_comparison.png}
%     \caption{Distribution of SET statistics for benign and backdoor inputs. Kernel density estimates show that, although the shifts vary across attacks, backdoor inputs remain consistently separable from benign inputs.}
%     \label{fig:set_density}
% \end{figure}

% To better understand why SET generalizes across diverse backdoor attacks, we visualize in Fig.~\ref{fig:set_density} the kernel density of the scalar statistic produced by SET for benign inputs and five representative backdoor attacks. Benign inputs form a relatively compact reference distribution, whereas backdoor inputs show attack-specific but still distinguishable deviations from that reference. These deviations do not follow a single common direction. Different attacks can produce different distribution shifts, yet the discrepancy from benign inputs remains systematic.

% This observation is consistent with our earlier CSRD analysis. Although different backdoor mechanisms perturb the conditional generation process in different ways, the scaling-induced response patterns captured by SET still differ from the benign response structure. This helps explain why SET retains strong detection performance across heterogeneous attacks, including more stealthy settings such as IBA.

\section{Experiments}
\subsection{Settings}
% \textbf{Attack Methods.} We consider five representative backdoor attack scenarios. Rickrolling~\cite{struppek2023rickrolling} and IBA~\cite{zhang2025iba} implant backdoors in the text encoder (i.e., CLIP~\cite{radford2021clip}). BadT2I~\cite{zhai2023badt2i} injects the backdoor into the UNet. VillanDiffusion~\cite{chou2023villandiffusion} fine-tunes the denoising component of Stable Diffusion while keeping the text encoder frozen. EvilEdit~\cite{wang2024eviledit} directly modifies the projection matrices in the cross-attention modules of the U-Net. Together, these attacks cover a wide range of trigger types and backdoor targets. More details about the attack methods are provided in the appendix.

\noindent\textbf{Attack Methods.} We consider five representative backdoor attack scenarios covering diverse attack mechanisms and trigger styles. Rickrolling~\cite{struppek2023rickrolling} and IBA~\cite{zhang2025iba} implant backdoors in the text encoder (i.e., CLIP~\cite{radford2021clip}). BadT2I~\cite{zhai2023badt2i} backdoors the model through data poisoning. VillanDiffusion~\cite{chou2023villandiffusion} fine-tunes the denoising component of Stable Diffusion while keeping the text encoder frozen. EvilEdit~\cite{wang2024eviledit} directly modifies the projection matrices in the cross-attention modules of the U-Net. Together, these attacks cover diverse backdoor injection pathways, target components, and trigger styles.

% \noindent\textbf{Baselines.} We consider four existing input-level backdoor defense baselines under the same setting: (i) $\text{T2IShield}_\text{FTT}$ and $\text{T2IShield}_\text{CDA}$~\cite{wang2024t2ishield}, (ii) DAA-I and DAA-S~\cite{wang2026daa}, (iii) NaviT2I~\cite{zhai2025navi}, and (iv) UFID~\cite{guan2025ufid}.

\noindent\textbf{Baselines.} We compare SET with four input-level backdoor defense baselines under the same setting, including methods based on cross-attention response analysis, i.e., $\text{T2IShield}_\text{FTT}$, $\text{T2IShield}_\text{CDA}$~\cite{wang2024t2ishield}, DAA-I, and DAA-S~\cite{wang2026daa}, a method based on neuron activation variation, i.e., NaviT2I~\cite{zhai2025navi}, and a black-box method based on generation consistency under input perturbations, i.e., UFID~\cite{guan2025ufid}.

\noindent\textbf{Datasets and Models.} To ensure a fair comparison, we use the MS-COCO dataset~\cite{lin2014coco} throughout. (1) For backdoor attacks that are not tied to specific input text, including Rickrolling, BadT2I, and Villan, we randomly sample 1,000 prompts from the MS-COCO val texts and inject triggers into half of them. (2) For EvilEdit, which targets specific objects in the text, we first randomly sample 500 prompts from the MS-COCO and additionally collect 500 prompts that contain the word "cat". We then inject triggers into these 500 samples, for example by replacing "cat" with "beautiful cat", to construct the attack inputs. We conduct our main experiments on Stable Diffusion v1.4~\cite{rombach2022ldm}, as it is widely used in prior studies on backdoor attacks and defenses.

\noindent\textbf{Metrics.} Following existing backdoor detection works~\cite{wang2024t2ishield,wang2026daa,zhai2025navi,guan2025ufid}, we adopt the area under the receiver operating characteristic curve (AUROC) as the primary metric for evaluating detection performance, since it avoids dependence on a specific threshold selection. We also report detection accuracy (ACC).

\subsection{Detection Results}
\begin{table}[t]
\centering
\caption{The performance (AUROC) against the mainstream T2I backdoor attacks on MS-COCO. We mark the best results in \textbf{bold} and the second-best results with underlines for comparison.}
\label{tab:auroc}
\setlength{\tabcolsep}{2pt}
\renewcommand{\arraystretch}{1.08}
\begin{tabular}{l|ccccc|c}
\toprule
Method & RickBKD & VillanBKD & BadT2I & EvilEdit & IBA & \textbf{Avg.} \\
\midrule
T2IShield$_{FTT}$ & \textbf{99.9} & 73.3 & 72.3 & 36.3 & 40.9 & 64.5 \\
T2IShield$_{CDA}$ & \textbf{99.9} & 96.1& \underline{99.8} & 45.0 & 5.5 & 69.3 \\
DAA-I & 84.0 & 74.8 & 87.6 & \underline{79.1} & 72.1 & 79.5 \\
DAA-S & 96.0& 82.0 & 97.3 & 76.6 & \underline{78.3} & \underline{86.0} \\
NaviT2I & \textbf{99.9}& 99.5& \textbf{99.9} & 70.8 & 54.9 & 85.0 \\
UFID & 37.4 & \textbf{99.9} & 28.7 & 49.0 & 13.7 & 45.7 \\
\rowcolor{gray!15}
\textit{\textbf{Ours}}& \underline{99.6}& \underline{99.6}& \textbf{99.9} & \textbf{83.4} & \textbf{92.9} & \textbf{95.1}\\
\bottomrule
\end{tabular}
\end{table}

\begin{table}[t]
\centering
\caption{Detection accuracy (ACC) against mainstream T2I backdoor attacks on MS-COCO. Key results are highlighted as in Tab.~\ref{tab:auroc}. For this binary classification task, even ``random guessing'' yields 50.0\%.}
\label{tab:acc}
\setlength{\tabcolsep}{2pt}
\renewcommand{\arraystretch}{1.08}
\begin{tabular}{l|ccccc|c}
\toprule
Method & RickBKD & VillanBKD & BadT2I & EvilEdit & IBA & \textbf{Avg.} \\
\midrule
T2IShield$_{FTT}$ & 85.7 & 67.6 & 64.4 & 42.4 & 40.1 & 60.0 \\
T2IShield$_{CDA}$ & \textbf{98.4} & 86.8 & 97.0 & 49.7 & 49.2 & 76.2 \\
DAA-I & 87.9 & 74.1 & 92.1 & \textbf{73.1} & 46.8 & 74.8 \\
DAA-S & 94.3 & 85.5 & \underline{98.0} & 65.0 & 48.6 & \underline{78.3}\\
NaviT2I & 88.2 & 93.2& 93.1 & 61.2 & \underline{52.1} & 77.6\\
UFID & 50.1 & \textbf{99.8} & 50.1 & 51.8 & 50.0 & 60.4 \\
\rowcolor{gray!15}
\textit{\textbf{Ours}}& \underline{97.9} & \underline{96.9}& \textbf{99.3} & \underline{65.1} & \textbf{64.9} & \textbf{84.8}\\
\bottomrule
\end{tabular}
\end{table}

As shown in Tab.~\ref{tab:auroc} and Tab.~\ref{tab:acc}, SET achieves the best average detection performance across five representative T2I backdoor attacks. This result suggests that active probing through cross-attention scaling can consistently extract informative signals for distinguishing benign inputs from backdoor inputs. Overall, SET also delivers more balanced performance across attacks, suggesting stronger generalization across attack types.

For individual attacks, in cases such as RickBKD, VillanBKD, and BadT2I, where abnormal cues are easier to expose, SET performs comparably to the strongest baseline and achieves the best results on several metrics. This suggests that when attack traces are more visible, SET retains recognition performance on par with strong existing baselines while avoiding large performance swings across attacks. Its advantages become more pronounced in harder cases such as EvilEdit and IBA. IBA is particularly challenging because it uses syntactic structure as the backdoor trigger, which greatly reduces the surface-level anomalies on which many existing methods rely. Even in this case, SET maintains stable discriminative ability. A similar trend is observed for EvilEdit, which directly modifies parameters related to cross-attention, where SET again achieves the best or near-best results. These findings suggest that as triggers move from explicit tokens to more concealed and natural forms, the signals used by existing methods become much weaker, while the cross-attention response divergence captured by SET remains relatively stable.

% For individual attacks, in cases such as RickBKD, VillanBKD, and BadT2I, where abnormal cues are easier to expose, SET performs comparably to the strongest baseline and achieves the best results on several metrics. This suggests that when attack traces are more visible, SET retains recognition performance on par with strong existing baselines while avoiding large performance swings across attacks.

% SET is even more effective in harder cases such as EvilEdit and IBA. IBA is particularly challenging because it uses syntactic structure as the backdoor trigger, which greatly reduces the surface-level anomalies that many existing methods rely on. Even in this case, SET still maintains stable discriminative ability. A similar trend appears for EvilEdit, which directly modifies parameters related to cross-attention, where SET again achieves the best or near-best results. These findings suggest that as triggers move from explicit tokens to more concealed and natural forms, the signals used by existing methods become much weaker, while the cross-attention response divergence captured by SET remains relatively stable.

By contrast, several baseline methods perform well only on a limited subset of attacks and degrade markedly under stealthier trigger forms. This pattern is consistent with their underlying detection mechanisms. Most existing input-level methods rely on abnormal tokens, local response imbalance, or instability in generated outputs that appears during standard inference, but such cues are often much harder to expose consistently under stealthy attacks. SET instead examines the overall response dynamics of the conditional injection pathway under controlled scaling and models them through the benign response space, which leads to more robust detection across a wider range of attack scenarios.

\begin{table*}[t]
\centering
\caption{Ablation study of SET under different configurations. The default configuration of SET (ours) uses the up and down layers, self-attention scaling, scaling factors $\{0.2, 0.3, 7, 10, 20\}$, 1,000 benign training samples, scaling at $in_V$, and the first five denoising steps. Each variant changes only one factor relative to this default setting.}
\label{tab:ablation_set}
\setlength{\tabcolsep}{3.0pt}
\renewcommand{\arraystretch}{1.08}
\small
\resizebox{\textwidth}{!}{%
\begin{tabular}{lccccccccccccccccccc}
\toprule
Metric
& SET (Ours)
& \multicolumn{4}{c}{Layer selection}
& Self-attn scaling
& \multicolumn{2}{c}{Scaling factors}
& \multicolumn{4}{c}{Benign training samples}
& \multicolumn{3}{c}{Scaling position}
& \multicolumn{4}{c}{Denoising step} \\
\cmidrule(lr){3-6}
\cmidrule(lr){7-7}
\cmidrule(lr){8-9}
\cmidrule(lr){10-13}
\cmidrule(lr){14-16}
\cmidrule(lr){17-20}
&
& all & up & mid & down
& No scaling
& F1& F2& 600 & 800 & 1200 & 1500
& $in_{\mathrm{noV}}$ & $out_V$ & $out_{\mathrm{noV}}$
& 1 & 3 & 7 & 9 \\
\midrule
AUROC$\uparrow$
& \textbf{95.1}
& 64.7 & 56.0 & 53.9 & 57.9
& 33.8
& 34.3 & 77.6
& 37.8 & 61.8 & 44.5 & 70.1
& 37.5 & 45.7 & 42.6
& 14.3 & 42.0 & 68.3 & 30.8 \\

ACC$\uparrow$
& \textbf{84.8}
& 63.3 & 59.4 & 50.0 & 68.8
& 51.8
& 59.5 & 75.3
& 54.9 & 63.7 & 62.1 & 74.7
& 57.6 & 53.3 & 58.4
& 49.5 & 57.2 & 71.6 & 49.5 \\
\bottomrule
\end{tabular}%
}
\end{table*}

\subsection{Ablation Studies}

We conduct six ablation studies to assess the main design choices in SET. Unless noted otherwise, the default configuration uses the up and down layers, applies both cross-attention and self-attention scaling, adopts the bidirectional scaling set $\{0.2, 0.3, 7, 10, 20\}$, uses 1{,}000 benign training samples, injects the perturbation at $in_V$, and probes the model over the first five denoising steps. As shown in Tab.~\ref{tab:ablation_set}, this configuration gives the best overall performance.

\noindent\textbf{Impact of Layer selection.}
Using only a single stage, or simply using all layers, consistently performs worse than the default setting. The best results come from combining the up and down blocks. This pattern indicates that the most informative scaling responses are concentrated in these two stages and that the information they provide is complementary. The middle block adds little beyond this, while aggregating all layers appears to introduce redundant or noisy responses that reduce the separation between benign and backdoor inputs. We therefore use the up and down layers in the final configuration.

\noindent\textbf{Impact of Self-Attention Scaling.}
We next study whether self-attention scaling remains useful when cross-attention scaling is already applied. Removing self-attention scaling causes a clear drop in performance, which suggests that probing only the text-conditioned pathway does not fully reveal the difference between benign and backdoor inputs. One possible reason is that cross-attention scaling changes how textual conditions are injected, whereas self-attention scaling also affects how these perturbed signals propagate through and interact within latent visual features. Using both therefore exposes a fuller response difference, so we retain self-attention scaling in the default setting.

% \noindent\textbf{Impact of the Scaling Factors.}
% We finally examine the scaling factors. Using only factors below 1 or only factors above 1 performs clearly worse than the default bidirectional set $\{0.2, 0.3, 7, 10, 20\}$. This indicates that benign and backdoor inputs respond differently under both weakened and strengthened attention, and that these two directions provide complementary diagnostic information. The mixed scale set therefore captures a richer response profile than a one-sided perturbation scheme, which is why we use it in the final configuration.
\noindent\textbf{Impact of the Scaling Factors.}
We further investigate the effect of the scaling-factor design. Specifically, \textbf{F1} denotes the one-sided scaling set below 1, i.e., $\{0.15, 0.2, 0.25, 0.3, 0.35\}$, and \textbf{F2} denotes the one-sided scaling set above 1, i.e., $\{5, 7, 10, 15, 20\}$. Compared with the default bidirectional setting $\{0.2, 0.3, 7, 10, 20\}$, using only \textbf{F1} or only \textbf{F2} yields consistently worse performance. This suggests that benign and backdoor inputs respond differently under both weakened and strengthened attention, and that these two perturbation directions provide complementary diagnostic signals. The bidirectional scaling design therefore captures a more informative response profile than either one-sided alternative and is adopted in the final configuration.

\noindent\textbf{Impact of the Number of Benign Training Samples.}
The size of the benign reference set also has a substantial effect on detection. The trend is not monotonic. With too few benign samples, the normal response space is not characterized well enough. Adding more clean samples does not always help, and can instead enlarge the benign region by introducing intra-class variability. Empirically, 1{,}000 benign samples provide the best balance among coverage, compactness, and efficiency, and we use this as the default choice.

\noindent\textbf{Impact of the Scaling Position.}
We also compare four ways of injecting the scaling factor based on the standard attention response $R=\mathrm{softmax}(S)V$. The default choice, $in_V$, uses $\mathrm{softmax}(\lambda S)V$, which scales the attention score before the softmax and then multiplies by $V$. The variant $in_{\mathrm{noV}}$ uses the same inside-softmax scaling but extracts the signal before multiplication with $V$, namely $\mathrm{softmax}(\lambda S)$. The variant $out_V$ scales the normalized attention matrix before multiplication with $V$, i.e., $\lambda\,\mathrm{softmax}(S)V$. Finally, $out_{\mathrm{noV}}$ scales the normalized attention matrix without the value projection, namely $\lambda\,\mathrm{softmax}(S)$. Among these variants, $in_V$ performs best by a clear margin. This suggests that the most effective probing strategy is to alter the competition among attention scores before normalization while preserving the full value-conditioned response. By comparison, scaling outside the softmax behaves like a uniform rescaling of magnitude, and removing $V$ discards content carried by the value vectors. We therefore adopt $in_V$ in the final design.

\noindent Impact of the Denoising Step.
The denoising step is critical. In the early denoising stage, the latent trajectory is shaped more directly by textual conditions, so cross-attention scaling tends to produce larger response differences between benign and backdoor inputs. Probing overly early steps is still suboptimal, however, because the extracted features are less stable and therefore less transferable across different backdoor attacks. Probing too late is also less effective, as the separation between benign and backdoor inputs gradually decreases in the later denoising stage, which in turn weakens their discriminative utility. In our experiments, probing the first five denoising steps provides the best balance between strong response divergence and cross-attack generalization. Although individual later steps can still provide useful signals, they are consistently less effective than this early multi-step setting. This observation motivates our use of the first five denoising steps as the default configuration.

% Overall, these ablations support the final design of SET. The best performance is obtained by combining complementary layers, applying both cross-attention and self-attention scaling, perturbing responses at $in_V$, probing at an intermediate denoising step, using a bidirectional scale set, and adopting a compact benign reference set of moderate size. Taken together, these choices allow SET to reveal the divergence between benign and backdoor responses more clearly while maintaining a compact and stable benign space for one-class detection.

\section{Conclusion}
In this paper, we studied input-level backdoor detection for text-to-image diffusion models under increasingly stealthy trigger settings, where existing defenses often become unreliable. From an active probing perspective, we introduced controlled scaling on cross-attention and identified a new phenomenon, \textbf{Cross-Attention Scaling Response Divergence (CSRD)}, which reveals stable and systematic differences between benign and backdoor inputs along the conditional generation pathway. Based on this insight, we proposed \textbf{SET}, a simple yet effective detection framework that constructs scaling-induced response-shift features and learns a compact benign response space from a small clean reference set. Extensive experiments across diverse attack methods, trigger types, and model settings showed that SET consistently outperforms existing baselines, with especially clear advantages under stealthy implicit-trigger attacks. These results suggest that controlled cross-attention scaling provides a robust and practical way to expose attack-agnostic discrepancies that are difficult to observe under standard inference.

\bibliographystyle{ACM-Reference-Format}
\bibliography{ref}

%
% If your work has an appendix, this is the place to put it.
\appendix
\section{Proof of Theorem 3.1}
\label{app}

In this appendix, we spell out the regularity conditions used in Theorem~3.1 and give the full proof. Throughout, we fix a denoising step $t$ and a cross-attention layer $\ell$, and write
\begin{equation}
\begin{aligned}
S(x) := S_{t,\ell}(x), \qquad
V(x) := V_{t,\ell}(x), \qquad \\
R^\lambda(x) := R^\lambda_{t,\ell}(x), \qquad
R(x) := R^1_{t,\ell}(x),
\end{aligned}
\end{equation}
and let $N := N_{t,\ell}$ denote the number of terms averaged in the MSE at the fixed step-layer pair $(t,\ell)$. Recall that
\begin{equation}
R^\lambda(x) = \operatorname{softmax}(\lambda S(x))V(x),
\end{equation}
where the softmax is applied row-wise.

The proof proceeds in three steps. We first state the local regularity conditions and define the first-order scaling sensitivity. We then derive the quadratic local expansion of the class-wise mean response-shift curve around $\lambda=1$. Finally, we show that unequal quadratic coefficients imply local separation between the benign and backdoor curves.

\subsection{Regularity Assumptions and Scaling Response Mechanism}\label{a.1}

Suppose $\widetilde{M}$ is the deployed model, which may be backdoored, and let $P_{\mathrm{ben}}$ and $P_{\mathrm{bd}}$ denote the benign and backdoor input distributions. For Theorem~3.1, the only model quantity we need is the class-conditional law of the scaling sensitivity of the cross-attention response at the fixed step-layer pair $(t,\ell)$.

We assume the following local regularity condition. There exists $\delta > 0$ such that, for $P_c$-almost every $x$ and each class $c \in \{\mathrm{ben},\mathrm{bd}\}$, the map
\begin{equation}
\lambda \mapsto R^\lambda(x)
\end{equation}
is twice continuously differentiable on the interval
\begin{equation}
I_\delta := (1-\delta,\,1+\delta).
\end{equation}
Moreover, there exist measurable envelope functions $M_1(x)$ and $M_2(x)$ such that, for every $\lambda \in I_\delta$,
\begin{equation}
\left\| \frac{\partial R^\lambda(x)}{\partial \lambda} \right\|_F \le M_1(x),
\qquad
\left\| \frac{\partial^2 R^\lambda(x)}{\partial \lambda^2} \right\|_F \le M_2(x),
\end{equation}
and
\begin{equation}
\mathbb{E}_{x \sim P_c}\!\left[ M_1(x)^2 + M_2(x)^2 \right] < \infty,
\qquad c \in \{\mathrm{ben},\mathrm{bd}\}.
\end{equation}

These assumptions are used only to justify two local analytic steps around $\lambda=1$: the second-order Taylor expansion of $R^\lambda(x)$ and the interchange of limit and expectation when passing from sample-wise convergence to class-wise convergence. In particular, the envelope bound on the first derivative implies
\begin{equation}
\left\|
\left.\frac{\partial R^\lambda(x)}{\partial\lambda}\right|_{\lambda=1}
\right\|_F
\le M_1(x)
\quad\text{for $P_c$-almost every }x,
\end{equation}
so the class-wise coefficient defined below is well defined and finite under the stated integrability condition.

Because the row-wise softmax map is smooth, the derivative with respect to $\lambda$ is well defined. For a fixed step-layer pair $(t,\ell)$ in a finite-dimensional attention block, $\lambda$ appears only through the smooth row-wise softmax map applied to $\lambda S(x)$.

Let
\begin{equation}
A^\lambda(x):=\operatorname{softmax}(\lambda S(x))
\end{equation}
be the row-wise normalized attention matrix, and let $a_i^\lambda(x)$ and $s_i(x)$ denote its $i$-th row and the $i$-th row of $S(x)$, respectively, where
\begin{equation}
a_i^\lambda(x):=\operatorname{softmax}(\lambda s_i(x)).
\end{equation}
Under the row-vector convention used throughout this appendix, we have
\begin{equation}
\frac{d}{d\lambda} a_i^\lambda(x)
=
s_i(x)
\Bigl(
\operatorname{Diag}(a_i^\lambda(x))
-
\bigl(a_i^\lambda(x)\bigr)^\top a_i^\lambda(x)
\Bigr).
\end{equation}

Because $V(x)$ does not depend on $\lambda$,
\begin{equation}
\frac{\partial R^\lambda(x)}{\partial \lambda}
=
\frac{\partial A^\lambda(x)}{\partial \lambda}V(x).
\end{equation}
For the $i$-th row, this gives
\begin{equation}
\frac{\partial R_i^\lambda(x)}{\partial\lambda}
=
s_i(x)
\Bigl(
\operatorname{Diag}(a_i^\lambda(x))
-
\bigl(a_i^\lambda(x)\bigr)^\top a_i^\lambda(x)
\Bigr)V(x).
\end{equation}

Hence the first-order scaling sensitivity
\begin{equation}
G(x) := \left. \frac{\partial R^\lambda(x)}{\partial \lambda} \right|_{\lambda=1}
\end{equation}
exists for $P_c$-almost every $x$. It captures the first-order response of the conditional pathway to scaling in the cross-attention scores.

Thus, for the purpose of Theorem~3.1, the effect of the backdoor at the fixed step-layer pair $(t,\ell)$ is characterized by the class-conditional second moments of $G(x)$. Define
\begin{equation}
\Gamma_c^{t,\ell}
:=
\mathbb{E}_{x \sim P_c}
\left[
\frac{1}{N}\|G(x)\|_F^2
\right],
\qquad c \in \{\mathrm{ben},\mathrm{bd}\}.
\end{equation}
By the derivative envelope at $\lambda=1$, we have $\|G(x)\|_F \le M_1(x)$ for $P_c$-almost every $x$, and therefore $\Gamma_c^{t,\ell}<\infty$ under the stated moment condition.

The expression for $G(x)$ indicates that the local scaling sensitivity is determined jointly by three factors: the attention competition structure encoded in the softmax Jacobian, the row-logit pattern $s_i(x)$, and the value-dependent semantics carried by $V(x)$. As a result, a backdoor can change $\Gamma_c^{t,\ell}$ by modifying how token competition is structured, how sharply attention concentrates, or how changes in attention are mapped by the value vectors into conditional responses.

This interpretation is clear in a two-token toy case. Consider a single attention row with two tokens and scalar values $v_1,v_2$. Let
\begin{equation}
s=(a,b),\qquad p(\lambda)=\operatorname{softmax}(\lambda s)=\bigl(p_1(\lambda),p_2(\lambda)\bigr).
\end{equation}
Then
\begin{equation}
R^\lambda = p_1(\lambda)v_1 + p_2(\lambda)v_2,
\end{equation}
Since
\begin{equation}
p_1'(\lambda)=p_1(\lambda)p_2(\lambda)(a-b),
\qquad
p_2'(\lambda)=-p_1(\lambda)p_2(\lambda)(a-b),
\end{equation}
we obtain
\begin{equation}
\frac{dR^\lambda}{d\lambda}
=
p_1(\lambda)p_2(\lambda)(a-b)(v_1-v_2),
\end{equation}
and hence
\begin{equation}
\left.\frac{dR^\lambda}{d\lambda}\right|_{\lambda=1}
=
p_1(1)p_2(1)(a-b)(v_1-v_2).
\end{equation}
Therefore, in this toy case, the leading quadratic coefficient of the sample-level response-shift statistic is
\begin{equation}
\frac{1}{N}\Bigl[p_1p_2(a-b)(v_1-v_2)\Bigr]^2.
\end{equation}
This makes clear that changes in attention competition or value contrast can alter the local scaling response, and therefore the class-dependent coefficient $\Gamma_c^{t,\ell}$.

% The expression for $G(x)$ indicates that the local scaling sensitivity is determined jointly by the attention competition structure encoded in the softmax Jacobian, the row-logit pattern $s_i(x)$, and the value-dependent semantics carried by $V(x)$. As a result, a backdoor can change $\Gamma_c^{t,\ell}$ by modifying the way token competition is arranged, the degree to which attention becomes concentrated, or the way value vectors convert attention changes into conditional responses.

% This interpretation becomes clear in a two-token toy example. Consider a single attention row with two tokens and scalar values $v_1,v_2$. Let
% \begin{equation}
% s=(a,b),\qquad p=\operatorname{softmax}(s)=(p_1,p_2).
% \end{equation}
% Then
% \begin{equation}
% R^\lambda = p_1^\lambda v_1 + p_2^\lambda v_2,
% \end{equation}
% and a direct calculation gives
% \begin{equation}
% \left.\frac{dR^\lambda}{d\lambda}\right|_{\lambda=1}
% =
% p_1p_2(a-b)(v_1-v_2).
% \end{equation}
% Therefore, in this toy case, the leading quadratic coefficient of the sample-level response-shift statistic is
% \begin{equation}
% \frac{1}{N}\Bigl[p_1p_2(a-b)(v_1-v_2)\Bigr]^2.
% \end{equation}
% This makes explicit that changes in attention competition or value contrast can alter the local scaling response, and therefore the class-dependent coefficient $\Gamma_c^{t,\ell}$.

No stronger structural decomposition of $S(x)$ or $V(x)$ is required for the theorem. In particular, Theorem~3.1 uses the explicit condition
\begin{equation}
\Gamma_{\mathrm{ben}}^{t,\ell} \ne \Gamma_{\mathrm{bd}}^{t,\ell}
\end{equation}
as a non-degeneracy condition at the fixed pair $(t,\ell)$, rather than deriving this inequality from additional model-specific assumptions in this appendix.

\subsection{Cross-Attention Scaling Response Divergence}\label{a.2}

Define the scaling-induced response shift
\begin{equation}
\Delta(x;\lambda) := R^\lambda(x) - R(x).
\end{equation}
By Taylor's theorem with integral remainder around $\lambda=1$, for every $x$ such that the regularity conditions in Sec.~\ref{a.1} hold,
\begin{equation}
R^\lambda(x)
=
R(x)
+
(\lambda-1)G(x)
+
\frac{(\lambda-1)^2}{2} H(x,\lambda),
\end{equation}
where
\begin{equation}
H(x,\lambda)
:=
\int_0^1 2(1-s)
\left.
\frac{\partial^2 R^\mu(x)}{\partial \mu^2}
\right|_{\mu = 1 + s(\lambda-1)}
\, ds.
\end{equation}
Therefore,
\begin{equation}
\Delta(x;\lambda)
=
(\lambda-1)
\left(
G(x) + \frac{\lambda-1}{2} H(x,\lambda)
\right).
\end{equation}
By the envelope assumption in Sec.~\ref{a.1}, for every $\mu \in I_\delta$,
\begin{equation}
\left\|\frac{\partial^2 R^\mu(x)}{\partial \mu^2}\right\|_F \le M_2(x).
\end{equation}
Using the integral representation of $H(x,\lambda)$ together with
\begin{equation}
\int_0^1 2(1-s)\,ds = 1,
\end{equation}
we obtain
\begin{equation}
\begin{aligned}
\|H(x,\lambda)\|_F
&\le
\int_0^1 2(1-s)
\left\|
\left.
\frac{\partial^2 R^\mu(x)}{\partial \mu^2}
\right|_{\mu = 1 + s(\lambda-1)}
\right\|_F ds \\
&\le M_2(x),
\qquad \lambda \in I_\delta.
\end{aligned}
\end{equation}

Now recall the sample-level response-shift statistic
\begin{equation}
D_{t,\ell}(x;\lambda)
:=
\mathrm{MSE}\!\bigl(R^\lambda_{t,\ell}(x), R^1_{t,\ell}(x)\bigr)
=
\frac{1}{N}\|\Delta(x;\lambda)\|_F^2.
\end{equation}
Here $N=N_{t,\ell}$ denotes the number of squared-difference terms averaged in the MSE at the selected step-layer pair $(t,\ell)$.

Substituting the expansion of $\Delta(x;\lambda)$ gives
\begin{equation}
D_{t,\ell}(x;\lambda)
=
\frac{(\lambda-1)^2}{N}
\left\|
G(x) + \frac{\lambda-1}{2} H(x,\lambda)
\right\|_F^2.
\end{equation}
Hence, for $P_c$-almost every $x$,
\begin{equation}
\lim_{\lambda \to 1}
\frac{D_{t,\ell}(x;\lambda)}{(\lambda-1)^2}
=
\frac{1}{N}\|G(x)\|_F^2,
\qquad c \in \{\mathrm{ben},\mathrm{bd}\}.
\end{equation}

We now pass from sample-wise convergence to class-wise convergence. Fix $c \in \{\mathrm{ben},\mathrm{bd}\}$. For every $\lambda$ with $|\lambda-1|<\delta$, the inequality
\begin{equation}
\|A+B\|_F^2 \le 2\|A\|_F^2 + 2\|B\|_F^2
\end{equation}
gives
\begin{equation}
\frac{D_{t,\ell}(x;\lambda)}{(\lambda-1)^2}
\le
\frac{2}{N}\|G(x)\|_F^2
+
\frac{(\lambda-1)^2}{2N}\|H(x,\lambda)\|_F^2.
\end{equation}
Using the bounds from Sec.~\ref{a.1}, we obtain
\begin{equation}
\frac{D_{t,\ell}(x;\lambda)}{(\lambda-1)^2}
\le
\frac{2}{N}M_1(x)^2
+
\frac{\delta^2}{2N}M_2(x)^2.
\end{equation}

Thus, for $P_c$-almost every $x$, the quantity
\begin{equation}
\frac{D_{t,\ell}(x;\lambda)}{(\lambda-1)^2}
\end{equation}
converges pointwise to
\begin{equation}
\frac{1}{N}\|G(x)\|_F^2
\qquad \text{as } \lambda \to 1,
\end{equation}
and is dominated by the integrable function
\begin{equation}
\frac{2}{N}M_1(x)^2+\frac{\delta^2}{2N}M_2(x)^2.
\end{equation}
Therefore, by the dominated convergence theorem,
\begin{equation}
\begin{aligned}
\lim_{\lambda \to 1}
\frac{\bar D_c^{t,\ell}(\lambda)}{(\lambda-1)^2}
&=
\lim_{\lambda \to 1}
\mathbb{E}_{x \sim P_c}
\left[
\frac{D_{t,\ell}(x;\lambda)}{(\lambda-1)^2}
\right] \\
&=
\mathbb{E}_{x \sim P_c}
\left[
\frac{1}{N}\|G(x)\|_F^2
\right] \\
&=
\Gamma_c^{t,\ell}.
\end{aligned}
\end{equation}
where
\begin{equation}
\bar D_c^{t,\ell}(\lambda)
:=
\mathbb{E}_{x \sim P_c}\!\left[D_{t,\ell}(x;\lambda)\right].
\end{equation}
Equivalently,
\begin{equation}
\bar D_c^{t,\ell}(\lambda)
=
\Gamma_c^{t,\ell}(\lambda-1)^2
+
o\!\left((\lambda-1)^2\right)
\qquad \text{as } \lambda \to 1.
\end{equation}
This proves the quadratic local expansion in Theorem~3.1.

\subsection{Completion of the Proof: Local Separation}

By Sec.~\ref{a.2}, for each class $c \in \{\mathrm{ben},\mathrm{bd}\}$,
\begin{equation}
\bar D_c^{t,\ell}(\lambda)
=
\Gamma_c^{t,\ell}(\lambda-1)^2
+
o\!\left((\lambda-1)^2\right)
\qquad \text{as } \lambda \to 1.
\end{equation}
Assume now that
\begin{equation}
\Gamma_{\mathrm{ben}}^{t,\ell} \ne \Gamma_{\mathrm{bd}}^{t,\ell}.
\end{equation}
Let
\begin{equation}
a := \Gamma_{\mathrm{bd}}^{t,\ell} - \Gamma_{\mathrm{ben}}^{t,\ell} \ne 0.
\end{equation}
Subtracting the two expansions gives
\begin{equation}
\bar D_{\mathrm{bd}}^{t,\ell}(\lambda)
-
\bar D_{\mathrm{ben}}^{t,\ell}(\lambda)
=
a(\lambda-1)^2
+
o\!\left((\lambda-1)^2\right).
\end{equation}
By the definition of the little-$o$ term, there exists $\varepsilon \in (0,\delta)$ such that
\begin{equation}
\left|
\bar D_{\mathrm{bd}}^{t,\ell}(\lambda)
-
\bar D_{\mathrm{ben}}^{t,\ell}(\lambda)
-
a(\lambda-1)^2
\right|
\le
\frac{|a|}{2}(\lambda-1)^2,
\qquad
0<|\lambda-1|<\varepsilon.
\end{equation}
Therefore, for every $\lambda$ satisfying $0<|\lambda-1|<\varepsilon$,
\begin{equation}
\left|
\bar D_{\mathrm{bd}}^{t,\ell}(\lambda)
-
\bar D_{\mathrm{ben}}^{t,\ell}(\lambda)
\right|
\ge
\frac{|a|}{2}(\lambda-1)^2
>
0.
\end{equation}
Hence
\begin{equation}
\bar D_{\mathrm{bd}}^{t,\ell}(\lambda)
\ne
\bar D_{\mathrm{ben}}^{t,\ell}(\lambda),
\qquad
0<|\lambda-1|<\varepsilon,
\end{equation}
which proves the local separation of the benign and backdoor class-wise mean response-shift curves in Theorem~3.1.

Finally, the sign of
\begin{equation}
\bar D_{\mathrm{bd}}^{t,\ell}(\lambda)
-
\bar D_{\mathrm{ben}}^{t,\ell}(\lambda)
\end{equation}
for $\lambda$ sufficiently close to $1$ is the sign of
\begin{equation}
a = \Gamma_{\mathrm{bd}}^{t,\ell} - \Gamma_{\mathrm{ben}}^{t,\ell}.
\end{equation}
The theorem places no restriction on the sign of $a$, so the divergence may be positive or negative depending on the attack. That is, the direction of
\begin{equation}
\bar D_{\mathrm{bd}}^{t,\ell}(\lambda)
-
\bar D_{\mathrm{ben}}^{t,\ell}(\lambda)
\end{equation}
depends on the attack and is not fixed in advance.

Theorem 3.1 is a local result around the baseline operating point $\lambda=1$: it explains why scaling is informative for a fixed step-layer pair $(t,\ell)$. SET builds on this local view by aggregating response-shift statistics over multiple selected tuples $(t,\ell,\lambda)$, so that informative coordinates can still form a stable class-dependent pattern in practice even when some coordinates are weak or degenerate. This interpretation is consistent with the ablation results in Table~3, which suggest that informative coordinates are concentrated in specific stages of the U-Net and in the early denoising phase. 

Accordingly, the bidirectional scaling factors and the additional self-attention scaling used in SET should be understood as practical means of strengthening and carrying forward the discrepancy identified by the local cross-attention analysis, rather than as evidence that the quadratic expansion remains quantitatively accurate far from $\lambda=1$.

% Theorem~3.1 is a local result around $\lambda=1$. Its purpose is to establish the existence of response divergence in a neighborhood of the unscaled operating point. The larger bidirectional scaling factors used in SET are chosen empirically to amplify this discrepancy in practice, rather than to suggest that the asymptotic expansion above remains quantitatively accurate far from $\lambda=1$.This interpretation is consistent with the ablation results in Table 3, where the bidirectional scaling design yields the strongest empirical performance.

% The theorem focuses on the core conditional mechanism in cross-attention because the detection feature itself is extracted from cross-attention responses. The additional self-attention scaling used in practice is best understood as an empirical amplification device. It helps the perturbation propagate through the latent feature space and makes the discrepancy easier to reveal, while the measured detection signal remains on the cross-attention pathway.

This completes the proof.

\end{document}